\documentstyle[aps,eqsecnum]{revtex}
\input epsf

\def\cF{{\cal F}}
\def\hP{{\hat P}}
\def\tPsi{{\tilde\Psi}}
\def\cW{{\cal W}}

\begin{document}
\twocolumn[\hsize\textwidth\columnwidth\hsize\csname
@twocolumnfalse\endcsname

\title{Quantum Black Holes from Quantum Collapse \\}
\author{Cenalo Vaz${}^{a}$ and Louis Witten$^{b}$\\}

\address{$^{a}$Faculdade de Ci\^encias e Tecnologia, \\
Universidade do Algarve, Faro, Portugal.\\
{\rm Email address: {\tt cvaz@ualg.pt}}}

\address{$^{b}$Department of Physics,\\
University of Cincinnati, Cincinnati, OH 45221-0011, USA.\\
{\rm Email address: {\tt witten@physics.uc.edu}}}
\maketitle
\thispagestyle{empty}

\begin{abstract}
The Schwarzschild black hole can be viewed as the special case of the marginally
bound Lema\^\i tre-Tolman-Bondi models of dust collapse which corresponds to a
constant mass function. We have presented a midi-superspace quantization of this
model for an arbitrary mass-function in a separate publication. In this communication
we show that our solution leads both to Bekenstein's area spectrum for black
holes as well as to the black hole entropy, which, in this context, is naturally
interpreted as the loss of information of the original matter distribution within
the collapsing dust cloud.
\end{abstract}

\vskip2pc]

\baselineskip10.8pt

\section{Introduction}

Due to the singularity theorems of Penrose, Geroch, and Hawking\cite{st} and the many examples
that have been found in various models of gravitational collapse \cite{colmod}, it has now come
to be generally accepted that the gravitational collapse of a sufficiently massive star will
lead eventually to the formation of a black hole.

In one particular class of models, the LeMa\^\i tre-Tolman-Bondi (LTB) \cite{ltb} models of
spherical, inhomogeneous dust collapse, the picture that emerges is simple and general enough
to be interesting. These models are all represented by a solution of Einstein's equations with
pressureless dust described by the stress tensor  $T_{\mu\nu} = \epsilon U_\mu
U_\nu$, where $\epsilon(t,r)$ is the energy density of the collapsing dust cloud. The solution is
characterized by two arbitrary functions of the coordinate $r$, the mass function, $F(r)$, and
the energy function, $f(r)$, the former being related to the initial energy density distribution
in the collapsing cloud and the latter to its initial velocity profile. The collapsing dust cloud
is imagined to be made up of shells, labeled by $r$, which successively collide at late times
into a central singularity. The classical solution is given by
\begin{eqnarray}
ds^2~~ &=&~~ d \tau^2~ -~ {{{R'}^2} \over {1 + f}} dr^2~ -~ R^2 d\Omega^2,\cr
\epsilon~~ &=&~~ {{F'}\over {R^2 R'}},~~ R^*~~ =~~ \pm \sqrt{f~ +~ {F \over R}},
\label{ltbsol}
\end{eqnarray}
where $R$ is the physical radius, $R'$ represents a derivative w.r.t. $r$ and $R^*$ represents
a derivative w.r.t. the dust proper time, $\tau$.

An important subclass of these solutions, in which the energy function is taken to be exactly
vanishing, describes the marginally bound models. In this case, using a freedom to scale the
initial distribution, the general solution of (\ref{ltbsol}) can be cast into the form
\begin{equation}
R^{3 \over 2}(\tau,r)~~ =~~ r^{3 \over 2}~ -~ {3 \over 2} \sqrt{F(r)}\tau.
\label{ltbrad}
\end{equation}
The epoch $R=0$ describes a physical singularity. Successive shells, labeled by $r$,
collapse into this singularity at proper times given by
\begin{equation}
\tau(r)~~ =~~ {{2r^{3 \over 2}} \over {3\sqrt{F(r)}}}.
\end{equation}
Various models, with remarkably different behaviors, are obtained from different choices of
the mass function, $F(r)$. Some models lead to the formation of a naked singularity for certain
initial data and of a black hole for other initial data. One particular model, with
$F(r) = 2M$ (a constant), does not describe dynamical collapse at all but rather a static
Schwarzschild black hole of mass $M$. This is most easily seen by performing the following
transformation from the Tolman-Bondi coordinates $(\tau, r)$ in (\ref{ltbsol}) to curvature
coordinates, $(T,R)$,
\begin{eqnarray}
R(\tau,r) &=& r\left[1 - {3 \over 2} \sqrt{{2M} \over
{r^3}}\tau\right]^{2/3}\cr T(\tau,r) &=& \tau - \sqrt{2M}\int dR{{\sqrt{R}} \over
{R-2M}},
\end{eqnarray}
to recover the usual static, Schwarzschild form of the metric. $T$ is the Schwarzschild
(Killing) time.

Black holes are extremely interesting objects. Although their temperature is exactly zero
Kelvin in classical general relativity, Bekenstein proposed \cite{bek1} that they have
a temperature and an entropy and should be treated as thermodynamic systems. Their temperature and
entropy are known from semi-classical arguments\cite{hawk1} to be fundamentally quantum mechanical,
yet the precise quantum origin of these properties remains shrouded in mystery despite the many
interesting proposals\cite{ent1} that have been made in recent years. Ultimately, black
holes are the classical end states of collapse and it should be of considerable interest to
understand these properties from a bona fide microcanonical ensemble of quantum states constructed
from a collapsing matter distribution. In this communication we will examine the quantum mechanics of
black holes from the vantage point of a midi-superspace quantization of the LTB models described above.
At least two problems are eliminated by this approach: (i) the proper time of the collapsing dust
enters naturally in the dynamical equation, eliminating the so-called ``problem of time'' that usually
plagues the canonical quantization of gravity and (ii) there are no operator ordering ambiguities,
the configuration space is flat and an inner product may be defined unambiguously on the Hilbert
space so that the usual probabilistic interpretation of quantum mechanics is retained.

\section{Quantization of Dust Collapse}

In a separate publication\cite{cv1} we have presented the canonical quantization of the marginally
bound models for a variety of mass functions, including those that admit both black holes and naked
singularities. After a series of canonical transformations, performed in the spirit of Kucha\v r's
remarkable reduction of static spherical geometries\cite{ku1}, it becomes possible to describe the
phase space by the dust proper time, $\tau(r)$, the physical radius, $R(r)$, the mass function,
$F(r)$ and their canonical momenta, $P_\tau(r), P_R(r)$ and $P_F(r)$. The collapse is reduced to
two classical constraints, one of which, the momentum constraint,
\begin{equation}
\tau' P_\tau + R' P_R + F' P_F \approx 0,
\label{diff}
\end{equation}
insures spatial
diffeomorphism invariance. The second, the Hamiltonian constraint,
\begin{equation}
(P_\tau + F'/2)^2 + \cF P_R^2 - {{{F'}^2} \over {4\cF}} \approx 0,
\label{dyn2}
\end{equation}
is responsible for (proper) time evolution (we have set $\cF = 1 - F/R$ in (\ref{dyn2})). The mass
function, $F(r)$, defines the particular collapse model being considered. This is best seen in
the Lagrangian description in which the general solutions of Einstein's equations are given by
(\ref{ltbsol}) and (\ref{ltbrad}). These solutions show that $F(r)$ is related to the proper
energy of the collapsing cloud. ($F'(r)$ is the energy per unit coordinate cell, $d^3x$, and is
required to be positive definite.) As a consequence of (\ref{ltbsol}), $F(r)$ determines
$\epsilon(r,\tau)$ and vice-versa, apart from a choice of scaling, $\epsilon(r,\tau)$ determines
$F(r)$ according to
\begin{equation}
F(r)~~ =~~ \int_0^r \epsilon(r',0) {r'}^2 dr',
\end{equation}
where we have used the scaling implicit in (\ref{ltbrad}). Both $F(r)$ and $\epsilon(r,\tau)$
are externally given functions that define the collapse. This comes about because the energy density,
$\epsilon(r,\tau)$, is a Lagrange multiplier, as seen from the form of the dust action,
\begin{equation}
S^d = -\frac{1}{8\pi} \int d^4 x \sqrt{-g}~ \epsilon(x)\left[g_{\alpha\beta} U^\alpha U^\beta + 1
\right],
\label{dust}
\end{equation}
that reproduces Einstein's equations. As a Lagrange multiplier, $\epsilon(r,\tau)$ enforces
``time-like dust'' {\it i.e.,} it requires the dust world-lines to be time-like geodesics. Thus
$F(r)$, like $\epsilon(r,\tau)$, can be chosen arbitrarily, but a choice specifies an initial energy
density distribution and thus a collapse model. In the quantum theory, $F(r)$ acts as a weight in
the potential term, which identifies the collapse model being quantized. Each model corresponds to
stellar collapse under certain initial conditions.

The DeWitt supermetric, considered on the effective configuration superspace, $(\tau,R)$, is
non-degenerate, can be read off directly from (\ref{dyn2}) and is found to be flat, positive
definite when $\cF > 0$ and indefinite when $\cF < 0$,
\begin{equation}
\gamma_{ab} = \left(\matrix{1&0\cr0&{1\over{\cF}}}\right).
\label{configmet}
\end{equation}
It happens that, for the spherical geometries being considered, $\cF = 0$ or $R = F$ describes
the apparent horizon while, for the black hole geometry, it is the event horizon. When $\cF \neq 0$,
the supermetric can be brought to a manifestly flat form by the coordinate transformation
\begin{equation}
R_* = \pm \int {{dR} \over {\sqrt{|\cF|}}}.
\label{rstar}
\end{equation}
The next step is to turn the canonical momenta into covariant functional differential operators
(covariant w.r.t. the supermetric, $\gamma_{ab}$) \cite{VD1}, according to
\begin{equation}
\hP_a = - i \nabla_a = -i \left({{\delta} \over {\delta X^a(r)}} + \Gamma_a\right),
\end{equation}
which act on a state functional, $\Psi[\tau,R,F]$. Then defining
\begin{equation}
\Psi[\tau,R,F] = e^{-{i\over 2}\int_0^\infty F'(r)\tau(r)  dr} \tPsi[\tau,R,F],
\end{equation}
the functional $\tPsi$ is seen to obey the (Wheeler-DeWitt) equation,
\begin{equation}
\left[\gamma^{ab} \nabla_a \nabla_b + {{{F'}^2} \over {4\cF}}\right] \tPsi[X,F] = 0,
\label{reducedeqn}
\end{equation}
which is similar in form to a Klein-Gordon equation for a scalar field with a potential. Because the
metric in (\ref{configmet}) is positive definite in the region $R > F$, corresponding, in
the collapse geometries being considered, to the region outside the horizon, and indefinite when $R < F$,
the functional equation is elliptic in the former and hyperbolic latter regions. It is convenient to
write this equation as a functional Schroedinger equation in ($\tau,R_*$) by taking the square-root
of (\ref{reducedeqn}) as follows,
\begin{equation}
i {{\delta \tPsi} \over {\delta \tau}} = {\hat h} \tPsi = \pm \sqrt{\mp{{\hat P}_*}^2 + {{{F'}^2}
\over {4\cF}}}~ \tPsi[\tau,R, F],
\label{dyn}
\end{equation}
where the negative sign within the square root refers to the region $R > F$ and the positive
sign to the region $R < F$. Thus the dust proper time may be identified as the time variable as
was originally proposed by Kucha\v r, Torre and Brown \cite{ku2},\cite{ku3}.

Any solution of (\ref{dyn}) must obey the diffeomorphism constraint
\begin{equation}
\left[\tau'{\delta \over {\delta \tau}} + R_*'{\delta \over {\delta R_*}} + F'
{\delta \over {\delta F}}\right] \tPsi[\tau,R,F] = 0
\label{diff}
\end{equation}
and we will first provide an ansatz for a solution to this equation: as long as $F'(r)$ is not
everywhere vanishing, take
\begin{equation}
\tPsi[\tau,R,F] = \exp\left[{1 \over 2} \int_0^\infty dr F'(r)\cW(\tau(r),R(r),F(r))\right],
\label{diffsol}
\end{equation}
where $\cW$ is an arbitrary, complex valued function of its arguments (and not their derivatives)
which is to be determined from (\ref{dyn}). To see that this is indeed a solution, we note that
the integrand in the exponent is clearly a spatial density because, while $\tau(r)$, $R(r)$ and
$F(r)$ are spatial scalars, $F'(r)$ is a density. It follows that the wave-functional in
(\ref{diffsol}) will obey the momentum constraint. Indeed, (\ref{diff}) simply requires that
$\cW$ admit no explicit dependence on $r$. Our ansatz (\ref{diffsol}) is not unique, of course,
but it is guided by a physical consideration, namely that in the classical theory $F'(r)$ is
related to the proper energy density of the collapsing shells of dust.

Together, equations (\ref{dyn}) and (\ref{diff}) define the quantum theory whose inner product is given
by the functional integral,
\begin{equation}
\langle \Psi_1, \Psi_2\rangle = \int {\cal D}R_* \Psi_1^\dagger \Psi_2 = \int \prod_r dR_*(r)
\Psi_1^\dagger \Psi_2.
\label{inprod}
\end{equation}
As pointed out in \cite{ku3}, the inner product defined in this manner ensures the hermiticity of the
momentum, ${\hat P}_*$, conjugate to $R_*$. The norm of a quantum state under this scalar product is
formally $\tau$ independent provided that ${\hat h}$ defined in (\ref{dyn}) is self-adjoint. It is
clear from (\ref{reducedeqn}) that this operator represents the proper energy of the shell labeled by
$r$, the total energy of the system being simply
\begin{equation}
H = \int_0^\infty dr h(r).
\end{equation}
It is self-adjoint only in the linear sub-space in which the operator $\mp{{\hat P}_*}^2 + {F'}^2/ 4\cF$
admits positive eigenvalues.

\section{Static Black Holes: Bound States and Mass Quantization of a Single Shell}

We now apply the above considerations to the static black hole. As we have argued, in the quantum
theory as in the classical theory, the mass function is externally specified and defines the collapse
model being considered. In order to specify it in a way that is most consistent with the classical
solution ($F = 2M$), we consider a mass function describing a single spherical shell of total mass $M$.
Spherical shells have been used often in studies of gravitational collapse and cosmology \cite{ss}.
The shell label is immaterial, but for concreteness let us suppose it is $r=0$. Then we define the
shell by the mass function
\begin{equation}
F(r) = 2M \theta (r),
\label{massfn}
\end{equation}
where $\theta(r)$ is the usual step function. Therefore,
\begin{equation}
F'(r) = 2M \delta(r),
\label{massfnder}
\end{equation}
which vanishes everywhere except on the shell itself where it contributes infinitely to the shell self-energy.
Equation (\ref{diffsol}) tells us that the problem of a single shell is essentially quantum {\it mechanical},
\begin{equation}
\tPsi[\tau,R,F] = \exp\left[M \cW(\tau,R,M)\right],
\label{sshell}
\end{equation}
where $\tau = \tau(0), R = R(0)$ and $F(0)=2M$ represent, respectively, the proper time, the radial
coordinate and the total mass of the single shell. The Schwarzschild black hole as a single shell has
but one degree of freedom and the functional equation (\ref{dyn}) turns into an ordinary
Schroedinger equation. $F'(r)$ in (\ref{massfnder}) contributes  $\delta(0)^2$ to the potential term in
(\ref{dyn}). We take this to be vanishing, in keeping with DeWitt's regularization\cite{dw1}.
This subtracts the shell's infinite self-energy. It should be remarked, however, that this
regularization scheme is not unique and that there have been attempts to define a quantum theory in
which $\delta(0) \neq 0$ (see \cite{dk}). There is, unfortunately, no general consensus on how the
coincidence limits in the Wheeler-DeWitt equation are to be treated and for the present we assume
DeWitt's regularization. We then have
\begin{equation}
i {{\partial \tPsi} \over {\partial \tau}} = \pm \sqrt{\mp{{\hat P}_*}^2}~ \tPsi = 0.
\label{ordschr}
\end{equation}
As is normal in the quantum theory, we assume that the wave function is $C^{(1)}$. We take
$R_*$ in (\ref{rstar}) to range over the entire real line extending, in the interior ($R<F$),
from $-\pi M$ to $+\pi M$,
\begin{eqnarray}
R_* &=& \pm \left[- \sqrt{R(2M-R)}\right.\cr
&+& \left. M \tan^{-1}\left[{{R-M} \over {\sqrt{R(2M-R)}}}\right]+ {{\pi M}\over 2}\right]
\label{rstarint}
\end{eqnarray}
(its magnitude is the radius of the wormhole throat) and, in the exterior ($R>F$), from $\pm
\pi M$ to $\pm \infty$,
\begin{eqnarray}
R_* &=& \pm \left[\sqrt{R(R-2M)}\right.\cr
&+& M \ln [R - M + \sqrt{R(R-2M)}]\cr
&-& \left. M\ln M + \pi M\right].
\label{rstarext}
\end{eqnarray}
The classical singularity occurs at $R_* = 0$.

The stationary states of the black hole are now quite easily described: in the interior
they are a superposition of ingoing and outgoing waves,
\begin{equation}
\tPsi[\tau,R_*] = A_\pm e^{-iM(\tau\pm R_*)}
\label{in}
\end{equation}
and in the exterior, they are exponentially decaying
\begin{eqnarray}
\tPsi[\tau,R_*] &=& B e^{-iM(\tau -i R_*)},~~~~~ R_* > \pi M\cr
&=& C e^{-iM(\tau + iR_*)},~~~~~ R_* < -\pi M.
\label{out}
\end{eqnarray}
Matching the wave function and its derivative at the horizon, one finds that the energy (mass) squared
of the black hole is quantized in half integer units,
\begin{equation}
M^2 = \left(n+{1 \over 2}\right) M_p^2,~~ \forall~~ n \in {\bf N} \cup \{0\}
\end{equation}
where $M_p$ is the Planck mass. This is the Bekenstein area spectrum\cite{bek1}. A similar result
was reported earlier by us in \cite{cv2} (although our construction in that work was not aided by
the collapse model we are considering here) with the difference that the energy of the ground state
was found to be exactly zero. On the contrary, the minimum shell mass is here found to be
$M_p/\sqrt{2}$. The reason for this discrepancy is that the wave-function in\cite{cv2} was taken
to be identically vanishing outside the horizon. This condition was too strong and in fact unnecessary.
It will be seen that the non-vanishing ground state energy is required for the interpretation of the
black hole entropy that follows.

It is worth seeing how the choice, (\ref{massfn}), reproduces the constraints for massive dust
shells that have been used by various authors in the past\cite{kw},\cite{flw}. These constraints
were expressed in the phase space constructed from the metric coefficients in
\begin{equation}
ds^2~ =~ N^2 dt^2 - L^2(dr - N^r dt)^2 - R^2 d\Omega^2
\label{gsmet}
\end{equation}
and their conjugate momenta, and were derived from a square-root form of the matter action. They
take the form,
\begin{eqnarray}
H^d~~ &=&~~ \delta(r-\hat{r}) \sqrt{m^2 + {{p^2} \over {{\hat L}^2}}}\cr
H^d_r~~ &=&~~ - p\delta(r-\hat{r}).
\label{dustconst2}
\end{eqnarray}
In this system, $(\tau,L,R,P_\tau,P_L,P_R)$, the dust constraints obtained from the quadratic
action (\ref{dust}), upon eliminating the energy density, are (see, for example, \cite{cv1} or
\cite{ku3})
\begin{eqnarray}
H^d~~ &=&~~ P_\tau \sqrt{1 + {{U_r^2} \over {L^2}}}\cr
H^d_r~~ &=&~~ - U_r P_\tau,
\label{dustconst1}
\end{eqnarray}
where $U_r = -\tau'$. It is easy to see that the constraints (\ref{dustconst1}) turn into
(\ref{dustconst2}) by taking $P_\tau = m \delta(r-\hat{r}) = F'(r)/2$, as we have above
($\hat{r}$ replaces ``$0$'' as the shell label). The dust supermomentum is $H^d_r = -P_\tau
U_r = m \tau'\delta(r-\hat{r}) = -p\delta (r-\hat{r})$, where we have set $p=mU_r$.  The
gravitational part of the constraints, of course, has the usual form.

\section{Static Black Holes: The Superposition of Shells and the Statistical Entropy}

Black holes are not generally formed by the collapse of a single, infinitesimally thin, shell of
matter but by many shells, progressively colliding into a central singularity. The static black
hole must be viewed as the final state of their collapse, {\it i.e.,} when all shells have collided
with the central singularity (see figure 1).
\vskip 2mm
\centerline{\epsfxsize=2.5in \epsfbox{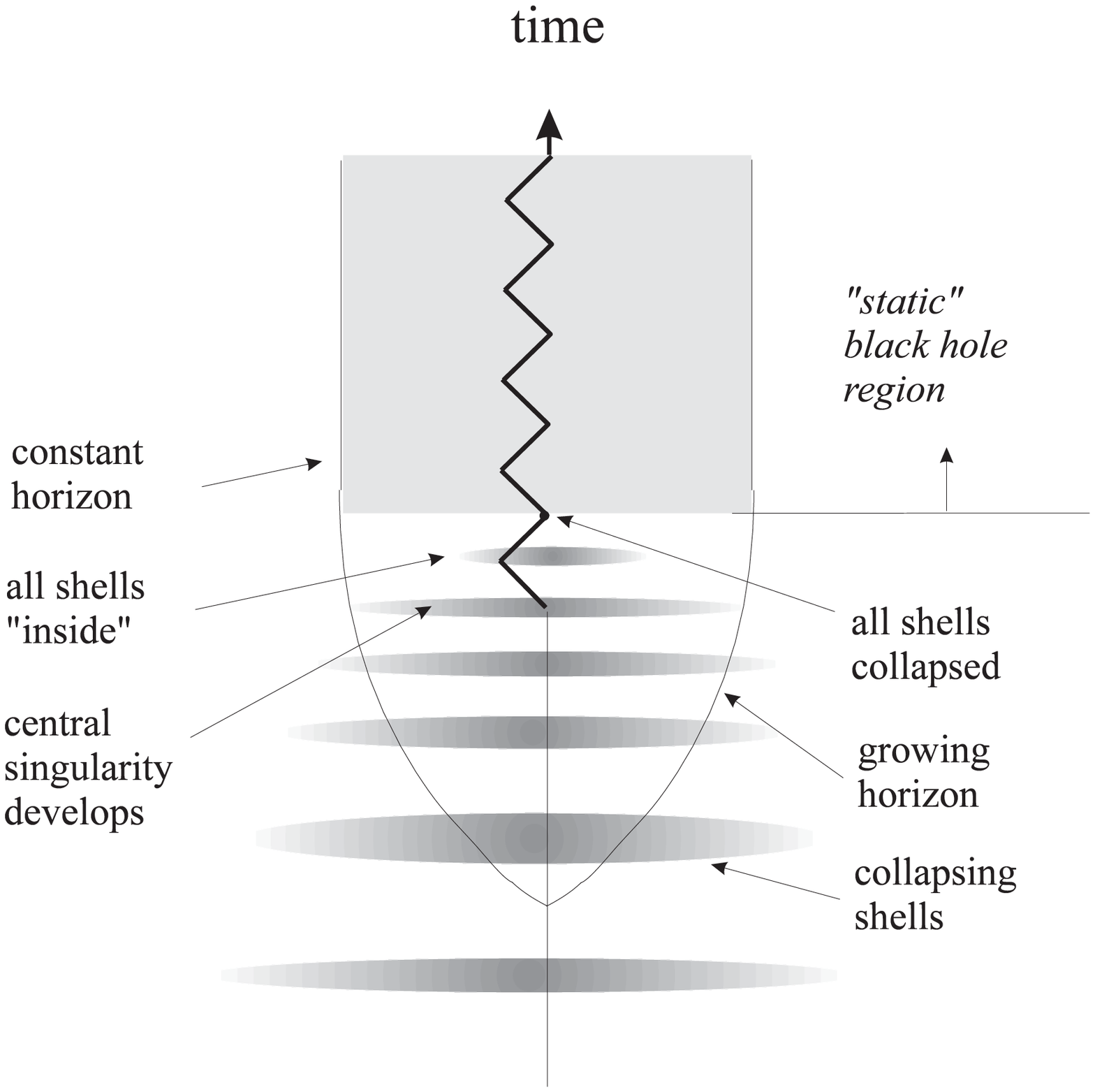}}
\centerline{\small {\bf Fig. 1}: Collapsing shells form a black hole}
\vskip 2mm
\noindent Let us therefore consider the description of $N$ shells. We take a
simple minded generalization of the mass function in (\ref{massfn}),
\begin{eqnarray}
F(r) &=& 2\sum_{j=1}^N \mu_j \theta(r-r_j)\cr
F'(r) &=& 2\sum_{j=1}^N \mu_j \delta(r-r_j).
\end{eqnarray}
Inserting $F'(r)$ into (\ref{diffsol}) shows that the wave-functional is now a product state
\cite{cv3},
\begin{equation}
\tPsi = \prod_{k=1}^N \tPsi_k(\tau, R_{*k},\mu_k)
\label{prodst}
\end{equation}
over $N$ wave-functions, one for each shell. The coordinate $R_{*k}$ is given by (\ref{rstarint})
and (\ref{rstarext}) and is determined not simply by the mass, $\mu_k$, of the shell (as it was
for a single shell) or the total mass, $M$, of all the shells, but by the total mass contained
{\it within} it,
\begin{equation}
2M_k = 2\sum_{j=1}^k \mu_j = \int_0^{r_k} dr F'(r) = F(r_k).
\end{equation}
Information about the gravitational interaction between shells is thus seen to be encoded in
$R_{*k}$, which is
\begin{eqnarray}
R_{*k} &=& \pm \left[- \sqrt{R_k(2M_k-R_k)}\right.\cr
&+& \left. M_k \tan^{-1}\left[{{R_k-M_k} \over {\sqrt{R_k(2M_k-R_k)}}}\right]+ {{\pi M_k} \over 2}
\right]
\end{eqnarray}
inside, and
\begin{eqnarray}
R_{*k} &=& \pm \left[\sqrt{R_k(R_k-2M_k)}\right.\cr
&+& M_k \ln [R_k - M_k + \sqrt{R_k(R_k-2M_k)}]\cr
&-& \left.M_k\ln M_k + \pi M_k\right]
\end{eqnarray}
outside. We see, not surprisingly, that the collapse of each shell is sensitive to the precise mass
distribution among the shells. Our system of $N$ shells has $N$ degrees of freedom and each wave-function
obeys the ordinary one dimensional Schroedinger equation in (\ref{ordschr}), with solutions given by
(\ref{in}) and ({\ref{out}), again eliminating the infinite shell self-interaction in the potential. A
straightforward application of the boundary conditions appropriate to each shell gives the following
quantization condition for the states of shell $k$,
\begin{equation}
\mu_k M_k = \left(n_k + {1 \over 2} \right) M_p^2.
\label{shellcond}
\end{equation}
These conditions, if applied recursively, show that the mass of shell $k$ is determined by $k$
quantum numbers. Thus the total mass depends on $N$ quantum numbers for a black hole formed out of
$N$ quantum shells.

We are now able to understand the origin of the black hole entropy in tangible terms. The appearance
of $N$ quantum numbers means a quantum black hole is not simply described by its total mass. Such a
description ignores the manner in which the mass is distributed among the shells. The entropy counts
the number of distributions for a given total mass.

The reasoning leading up to (\ref{shellcond}) cannot be completely extended to the static
black hole region. This is because it relies on a certain classical ordering of the shells, {\it i.e.,} shell one
is ``inside'' shell two, which is ``inside'' shell three, etc. But this ordering makes sense only as
long as the shells have not collided into the central singularity ($R=0$). Because, in the static region
all the shells have collapsed to the same physical point, all information of the original spatial
mass distribution is completely lost and we know only that there is {\it one} horizon (of physical
radius $R=2M$) for {\it all} the shells. Consequently, the mass condition in (\ref{shellcond})
should be replaced by the simpler relation
\begin{equation}
\mu_k M = \left(n_k + {1 \over 2} \right) M_p^2
\end{equation}
where $M$ is the {\it total} mass of the hole. Again, the total mass (squared) of the hole continues
quantized,
\begin{equation}
M = \sum_{j=1}^N \mu_j = {1 \over M} \sum_j \left(n_j + {1 \over 2} \right) M_p^2,
\label{massq}
\end{equation}
but now in integer as well as half integer units.

The problem of counting the number of distributions will be recognized as one from elementary statistical
mechanics texts. It is precisely the problem of $N$ simple harmonic oscillators into which one wishes to
distribute a total number, say $Q$, of quanta. (Alternatively, ask for the number of ways in which $N$
integers may be added to give another integer, $Q$.) If we knew the number of shells that went into the
black hole's making, the answer would be
\begin{equation}
\Omega = {{(N+Q-1)!}\over {(N-1)! Q!}}.
\end{equation}
In this case we have $Q=(M/M_p)^2-N/2$, from (\ref{massq}). The statistical entropy would then be
(exactly)
\begin{equation}
S = k\ln\left[{{(N/2+(M/M_p)^2-1)!}\over {(N-1)! [(M/M_p)^2 -N/2]!}}\right].
\end{equation}
This result depends on the number of shells that have collapsed to form the black hole. However we
do not know the number of shells that formed the black hole because the macrostate is defined only by the black
hole mass, so this number should be independently determined. For an eternal black hole in equilibrium it is
natural to maximize the entropy with respect to $N$. When both $N$ and $M/M_p$ are large, one readily finds
\begin{equation}
N \approx {2\over {\sqrt{5}}} \left({M \over {M_p}}\right)^2,
\label{nval}
\end{equation}
giving, to leading order,
\begin{equation}
S \approx k \ln p \left({M \over {M_p}}\right)^2
\end{equation}
where
\begin{equation}
p = {{\left(1+{1 \over {\sqrt{5}}}\right)^{\left(1+{1 \over {\sqrt{5}}}\right)}}\over
{\left({2 \over {\sqrt{5}}}\right)^{2 \over {\sqrt{5}}}
\left(1-{1 \over {\sqrt{5}}}\right)^{\left(1-{1 \over {\sqrt{5}}}\right)}}} \approx 2.618.
\end{equation}
If we now insert (\ref{nval}) into (\ref{massq}) we find that the total mass of the black hole is
quantized according to the relation $M^2 \approx \nu M_p^2/(1-1/\sqrt{5})$ for $\nu \in {\bf N}$, so
that the entropy can be given as $S \approx \nu k \ln p/(1-1/\sqrt{5})$. The average mass square of
each quantum shell is $\sqrt{5}M_p^2/2 \sim M_p^2$, therefore each shell may be thought of as
contributing about one Planck mass to the total mass of the black hole. In other words the shells,
on average, are all virtually in their ground states.

\section{Conclusions}

We have shown how the black hole area quantization and entropy can be understood in terms of the collapse
of shells of matter. The entropy encodes the loss of information of the matter distribution among the shells
in the final state. Shell states do not in general vanish at the classical singularity,
so the classical singularity is not prohibited by the quantum theory. However, it is interesting
that the wave function at the singularity either vanishes or has vanishing derivative.

What are the solutions when the matter distribution is continuous? Various continuum models lead to the
formation of both naked singularities as well as black holes \cite{colmod}, therefore the quantization
of the continuum models may be expected to shed some light on Hawking radiation and the Cosmic Censorship
Hypothesis. A solution for a general differentiable mass function was given in\cite{cv1}. However, a
realistic collapse would involve one or more regions of differing mass functions and while the mass
function must be continuous across the boundary between the regions, it may not necessarily be
differentiable. The wave-functional must be appropriately matched, {\it i.e.,} required to be both
continuous and differentiable at every such boundary. Specific models will be described elsewhere.
\vskip 5mm

\centerline{\bf ACKNOWLEDGEMENTS}
\vskip 3mm
\noindent We acknowledge the partial support of FCT, Portugal, under contract number POCTI/32694/FIS/2000.
L.W. was supported in part by the Department of Energy, USA, under contract Number DOE-FG02-84ER40153.


\begin{thebibliography}{99}
\bibitem{st}R. Penrose,  Phys. Rev. Letts. {\bf 14} (1965) 57; R. Geroch, Ann. Phys. {\bf 48} (1968) 526;
S. Hawking and R. Penrose, Proc. R. Soc. Lon. {\bf A314} (1969) 529.
\bibitem{colmod}see for example, P.S. Joshi, {\it Global Aspects in Gravitation and Cosmology}, Clarendon
Press, Oxford, (1993).
\bibitem{ltb}G. LeMa\^\i tre, Ann. Soc. Sci. Bruxelles I, {\bf A53} (1933) 51; R.C. Tolman,
Proc. Natl. Acad. Sci., USA {\bf 20} (1934) 410; H. Bondi, Mon. Not. Astron. Soc. {\bf 107}
(1947) 343.
\bibitem{bek1}J. D. Bekenstein, Ph. D. Thesis, Princeton University, April 1972; {\it ibid},
Lett. Nuovo Cimento {\bf 4} (1972) 737; Phys. Rev. {\bf D7} (1973) 2333; Phys. Rev. {\bf D9}
(1974); Phys. Lett. {\bf B360} (1995) 7; Phys. Rev. Lett. {\bf 70} (1993) 3680.
\bibitem{hawk1}S. W. Hawking, Commun. Math. Phys. {\bf 43} (1975) 199.
\bibitem{ent1}K. V. Krasnov, Gen. Rel. Grav. 30 (1998) 53; A. Ashtekar, J. Baez, A. Corichi, K Krasnov,
Phys. Rev. Letts. {\bf 80} (1998) 904; A. Ashtekar, J.Lewandowski, Class. Quant. Grav. {\bf 14} (1997)
A55; H. A. Kastrup, Phys. Lett. {\bf B385} (1996) 75; Phys. Letts. {\bf B413} (1997) 267; Phys. Letts.
{\bf B419} (1998) 40; A. W. Peet, Class. Quant. Grav. {\bf 15} (1998) 3291; K. Sfetsos, K. Skenderis,
Nucl. Phys. {\bf B517} (1998) 179; A. Strominger and C. Vafa, Phys. Lett. {\bf B379} (1996) 99;  C. O.
Lousto, Phys. Rev. {\bf D51} (1995) 1733; M. Maggiore, Nucl. Phys. {\bf B429}  (1994) 205;  Ya. I. Kogan,
JETP Lett. {\bf 44} (1986) 267; A. Strominger, JHEP 9802 (1998) 009; J.M. Maldacena and A. Strominger,
JHEP 9802 (1998) 014; J.M. Maldacena, A. Strominger and E. Witten, JHEP 9712 (1997) 002; G. Horowitz,
D. Lowe and J.M. Maldacena, Phys. Rev. Lett. {\bf 77} (1996) 430-433.
\bibitem{cv1}Cenalo Vaz, Louis Witten and T.P. Singh, Phys. Rev. {\bf D63} (2001) 104020.
\bibitem{ku1}K. V. Kucha\v r, Phys. Rev. {\bf D50} (1994) 3961.
\bibitem{VD1}G.A. Vilkovisky, Nucl. Phys. {\bf B234} (1984) 125; in ``Quantum Theory of Gravity'', ed.
S.M. Christensen (Adam Hilger, Bristol, 1984); A.O. Barvinsky and G.A. Vilkovisky, Phys. Rep. {\bf 119}
(1985) 9; B.S. DeWitt, in ``Quantum Field Theory and Quantum Statistics: Essays in Honour of
the Sixtieth Birthday of E.S. Fradkin'', eds. I.A. Batalin, C.J. Isham and G.A. Vilkovisky, (Adam Hilger,
Bristol, 1987).
\bibitem{ku2}K. Kucha\v r and C.G. Torre, Phys. Rev. {\bf D43} (1991) 419.
\bibitem{ku3}J.D. Brown and K.V. Kucha\v r, Phys. Rev. {\bf D51} (1995) 5600.
\bibitem{ss}see, for example, P. H\'aj\'\i \v cek, gr-qc/0007005; F. Larsen, F. Wilczek, Phys. Letts.
{\bf B375} (1996) 37; P. H\'aj\'\i \v cek, B. S. Kay and K. V. Kucha\v r, Phys. Rev. {\bf D46} (1992)
5439; E. Farhi, A. Guth and J. Guven, Nucl. Phys. {\bf B339} (1990) 417.
\bibitem{dw1}B.S. DeWitt, Phys. Rev. {\bf 160} (1967) 1113.
\bibitem{dk}J.-G. Demers and C. Kiefer, Phys. Rev. {\bf D53} (1996) 7050.
\bibitem{cv2}Cenalo Vaz and Louis Witten, Phys. Rev. {\bf D60} (1999) 024009.
\bibitem{kw}Per Kraus and F. Wilczek,  Nucl. Phys.  {\bf B433} (1995) 403; {\it ibid}, Nucl. Phys.
{\bf B437} (1995) 231.
\bibitem{flw}J.L. Friedman, J. Louko and S. Winters-Hilt, Phys. Rev. {\bf D57} (1998) 2279.
\bibitem{cv3}Cenalo Vaz, Phys. Rev. {\bf D61} (2000) 064017.
\end{thebibliography}
\end{document}